\documentclass[twocolumn,showpacs,preprintnumbers,amsmath,amssymb]{revtex4}

\usepackage{graphicx}
\usepackage{dcolumn}
\usepackage{bm}

\begin{document}

\preprint{}

\title{Long Time Tail of the Velocity Autocorrelation Function in a Two-Dimensional Moderately Dense Hard Disk Fluid}

\author{Masaharu Isobe}
\email{isobe@nitech.ac.jp}

\affiliation{
Graduate School of Engineering, Nagoya Institute of Technology, Nagoya 466-8555, Japan}

\date{\today}

\begin{abstract}
Alder and Wainwright discovered the slow power decay $\sim t^{-d/2}$ ($d$:dimension) of the velocity autocorrelation function in moderately dense hard sphere fluids using the event-driven molecular dynamics simulations.
In the two-dimensional case, the diffusion coefficient derived using the time correlation expression in linear response theory shows logarithmic divergence, which is called the ``2D long-time-tail problem''.
We revisited this problem to perform a large-scale, long-time simulation with one million hard disks using a modern efficient algorithm and found that the decay of the long tail in moderately dense fluids is slightly faster than the power decay ($\sim 1/t$). We also compared our numerical data with the prediction of the self-consistent mode-coupling theory in the long time limit ($\sim 1/(t\sqrt{\ln{t}})$).
\end{abstract}

\pacs{05.10.-a, 05.20.Jj, 61.20.Lc}

\maketitle


More than thirty-five years ago, Alder and Wainwright discovered the slow power decay $\sim t^{-d/2}$ ($d$:dimension) of the velocity autocorrelation function (VACF) in moderately dense hard sphere fluids using an event-driven molecular dynamics (EDMD) simulation~\cite{alder_1970}.
In the two-dimensional (2D) case ($d=2$), the transport coefficient (i.e., the diffusion constant $D$) derived using the time-correlation function in linear response theory, the so-called Green-Kubo expression~\cite{kubo_1991}, shows logarithmic divergence.
This means that conventional hydrodynamics do not exist, and this is called the ``2D long-time-tail problem''~\cite{pomeau_1975, dorfman_1977}.
This discovery has greatly influenced the development of non-equilibrium statistical physics and liquid states; several theories based on a kinetic approach~\cite{dorfman_1970} and mode coupling theory (MCT)~\cite{ernst_1970, kawasaki_1970} have been constructed.
From the numerical perspective, since a longitudinal sound wave propagates at the speed of sound $c_s$ in a system with periodic boundary conditions (PBCs), the maximum correlation time for obtaining the true VACF in a numerical simulation is limited by the system size, i.e., the particle number, $N$.
Therefore, we must perform a large-scale simulation to explore the exact form of the long time tail, which is not yet fully understood~\cite{petrosky_1999}.

Alder and Wainwright~\cite{alder_1970} obtained the VACFs in a hard disk system with about one thousand hard disks and found that long power decay occurred instead of ordinary exponential decay.
Erpenbeck and Wood also examined this system in a direct simulation with several thousand hard disks~\cite{erpenbeck_1982}.
In the 1990s, Frenkel and Ernst, Hoef and Frenkel, Naitoh et al., and Lowe and Frenkel investigated a systematic large-scale simulation using a lattice gas cellular automaton model and conducted theoretical analyses~\cite{frenkel_1989, hoef_1990, naitoh_1990, lowe_1995}.
Since their system is discrete in space, time, and velocity, the long time tail obtained from very efficient simulations can be compared with the predictions of the self-consistent mode coupling theory (SCMCT)~\cite{kawasaki_1971, wainwright_1971}, in which they showed a clear evidence of a good agreement between them.
It has been very difficult to investigate the long tail using a direct hard disk molecular dynamics simulation using an event-driven scheme for several reasons.
First, since a sound wave propagates through PBCs and the true VACF is disturbed, a large system with many particles is needed.
Second, to discuss the tail of a long time correlation, we must perform a long time simulation for each parameter.
Finally,  to investigate the functional form of the tails in the VACF accurately, we need to average numerous statistical samples (ensembles) of independent physical properties such as velocities.
Recently, the development of the computer and sophisticated modern algorithms ~\cite{rapaport_1980, marin_1993, isobe_1999} have enabled us to perform massive molecular dynamic simulations of a hard sphere system.

In this study, we revisit the ``2D long-time-tail problem'' and perform a large-scale, long-time, statistically accurate, systematic EDMD simulation with one million hard disks using a fast modern algorithm~\cite{isobe_1999}.
We found that the decay of the VACF in moderately dense fluids is slightly faster than ($\sim 1/t$), which seems to agree with the prediction of the SCMCT in the long time limit ($\sim 1/(t\sqrt{\ln{t}})$).


The Green-Kubo expression for the diffusion coefficient $D$ is described by

\begin{equation}
D \sim \int_{0}^{\infty} \langle v_x(t)v_x(0) \rangle dt
\label{eqn:1}
\end{equation}

\noindent
where $v_x(t)$ is the velocity of the tagged particle at time $t$ and $\langle \cdot \rangle$ indicates ensemble averages.
In conventional kinetic theory, the time correlation function $\langle v_x(t)v_x(0) \rangle$ decays exponentially as $\langle v_x(t)v_x(0) \rangle \sim \exp{(-t/\tau_E)}$, where $\tau_E$ is the relaxation time.
However, Alder and Wainwright~\cite{alder_1970} discovered the long time tail as $\langle v_x(t)v_x(0) \rangle \sim \left( t/\tau_E \right)^{-d/2}$.
In the case $d=2$, the long time tail derives the logarithmic divergence as $D \sim \ln{(\infty)} + const.$, in which the hydrodynamics would break down.


Our system consists of about one million elastic 2D hard disks ($N=1024^2$) placed in a $L_x \times L_y$ square box with PBCs.
The basic units in this system are mass $m$, disk diameter $a_0$, and energy $k_BT$.
The time unit can be described as $\sqrt{ma_0^2/k_BT}$.
Initially, the simulation systems for each packing fraction $\nu = N \pi (a_0/2)^2/L_xL_y$,
are prepared as the equilibrium state in a sufficiently long preliminary run, in which the density is uniform and the disk velocities have a Maxwell-Boltzmann distribution.
The system evolves through collisions, with up to $10^5$ collisions per particle, using a modern fast algorithm based on event-driven molecular dynamics~\cite{isobe_1999}.
To obtain accurate VACFs, we average samples using the particle number $N$ and velocities in two directions ($v_x,v_y$).
In this system, there are several time scales; the mean free time $t_0$, 
and the typical relaxation time $\tau_E$ at which the decay gradually 
changes from the exponential to the power form.
The fast longitudinal sound wave that results from a particle collision propagates with a sound velocity of $c_s$ through the ``artificial'' PBCs.
This artificial effect appears in the VACF around time $t_{max} \sim (L/2)/c_s$.
If we want to investigate pure VACFs at the thermodynamic limit, it is important to evaluate $t_{max}$ (i.e., the sound velocity $c_s$).
In all of the VACF figures in this letter, the time is scaled using the mean free time $t_0$ and the VACFs are normalized using the square of the initial velocity $\langle v(0)^2 \rangle$.

Here, we summarize the theoretical estimation of the mean free time $t_0$ and the sound velocity $c_s$ in a hard disk fluid.
We refer to 2D Enskog theory~\cite{gaspard_2004}
In Eq.(89) of Ref.~\cite{gaspard_2004}, the mean free path $l$ as a function of number density $n$ is described by 

\begin{equation}
l(y) \simeq \frac{1}{2\sqrt{2}n a_0 Y(y)} = \frac{\pi a_0}{8\sqrt{2} y Y(y)},
\end{equation}

\noindent
where  $y=(\pi/4) n a_0^2$ and $Y(y)=(1-(7/16)y)/(1-y)^2$ (Enskog factor).
The non-dimensional mean free path $l^*(y)(=l(y)/a_0)$ becomes

\begin{equation}
l^*(y) \simeq \frac{\pi}{8\sqrt{2}y Y(y)}.
\end{equation}

\noindent
Therefore, the non-dimensional mean free time $t_0^*$ can be estimated easily using $l^*(y)$ and the mean particle velocity.
The non-dimensional sound velocity $c^*_s$ in units of ($a_0,m, k_BT$) in a hard disk fluid for each density is also estimated using Enskog theory, and can be described as

\begin{equation}
c^*_s=\sqrt{f+f^2+y\frac{df}{dy}}.
\end{equation}

\noindent
where $f(y)=1+2yY(y)$ and $df/dy=2(1+y/8)/(1-y)^3$.
We confirmed that these theoretical values are in quite good agreement with numerical simulations for a wide range of densities, except around the solid-fluid transition density (the so-called Alder transition $\nu \sim 0.70$).


\begin{figure}
\includegraphics[scale=0.4]{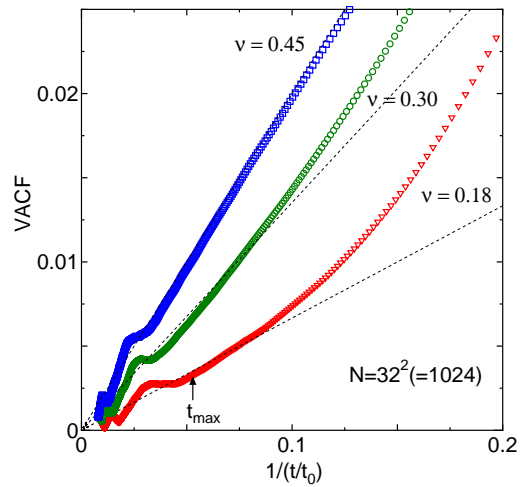}
\caption{
The VACFs for a typical packing fraction in terms of inverse time are shown.
The particle number and packing fractions are fixed at $N=1024$ and $\nu=0.18,0.30,0.45$, respectively.
With these parameters, this figure corresponds to Fig.~3 in Ref.~\cite{alder_1970}.
}
\end{figure}

Figure 1 shows the VACFs in a system with $32^2 (=1024)$ hard disks, which is almost the same size as Fig.~3 of Ref.~\cite{alder_1970}, in which the maximum system size was $N=986$.
The packing fractions $\nu$ ($= 0.45, 0.30, 0.18$) in Fig.~1 correspond to $A/A_0$ ($=2, 3, 5$) in Ref.~\cite{alder_1970}, where $A/A_0$ means the area $A$ of the system divided by the area $A_0$ at the close packing density.
(The relation between $\nu$ and $A/A_0$ is $\nu=\pi/(2\sqrt{3}(A/A_0))$.)
We found that the curves obtained using our numerical simulation are quite smooth and consistent with Fig.~3 in Ref.~\cite{alder_1970}.
In this small system, it is difficult to discuss the exact functional form of the tail and its exponents, since the region between $\tau_E$ and $t_{max}$ is quite narrow.


\begin{figure}
\includegraphics[scale=0.4]{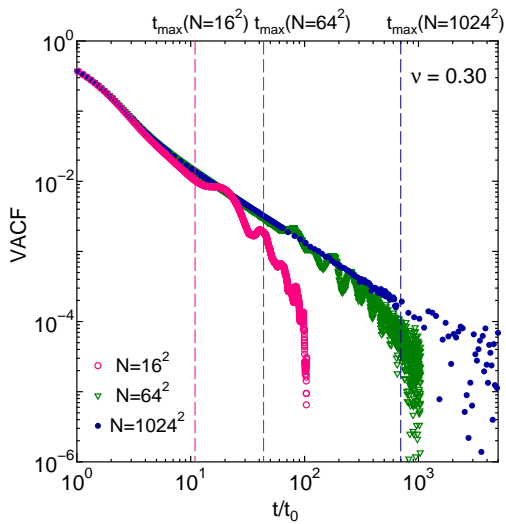}
\caption{
The system-size dependence of the VACFs on the packing fraction $\nu=0.30$.
}
\end{figure} 

To clarify the effect of sound wave propagation due to PBCs, we investigated the system-size dependence of the VACF by changing the particle number systematically.
Figure 2 shows VACFs of various sizes at a fixed packing fraction ($\nu=0.30$).
We found clear evidence of artificial disturbance of the sound wave by PBCs in VACFs, in which they deviate from the pure VACF around time $t_{max}$.
For instance, in the case $N=16^2 (=256)$, it deviates from the pure VACF at around $t/t_0 \sim 10$, which corresponds to $t_{max}$ for the system $N=16^2 (=256)$.
The same results were obtained at other packing fractions.
These facts confirm the validity of the sound velocity $c_s$ estimated using Enskog theory in a hard disk fluid.
The maximum size of our system is about one million particles ($N=1024^2$), for which $t_{max}$ is estimated as $t/t_0 \sim 698$ at $\nu=0.30$.
Our results can be used to discuss the pure VACF of the thermodynamic limit at more than $t/t_0 \sim 500$.
In the following, we analyse only the data of the VACFs with the maximum system size ($N=1024^2 = 1048576$).

Under the assumption that the decay of the long tail has the power form, as indicated by Ref.~\cite{alder_1970}, we first fitted the numerical data between $\tau_E$ and $t_{max}$ with a function in the form $\sim (t/t_0)^\alpha$ by changing the parameter $\alpha$.
We found that the exponent $\alpha$ obviously deviated from $-1$, and had values of $\alpha \sim -1.08, -1.06$, and $-1.03$ for $\nu=0.18,0.30$,and $0.45$ respectively.
The numerical data of the long time tails in a moderately dense system obtained from our simulation seem to decay faster than the power form ($\sim 1/t$).


\begin{figure}
\includegraphics[scale=0.4]{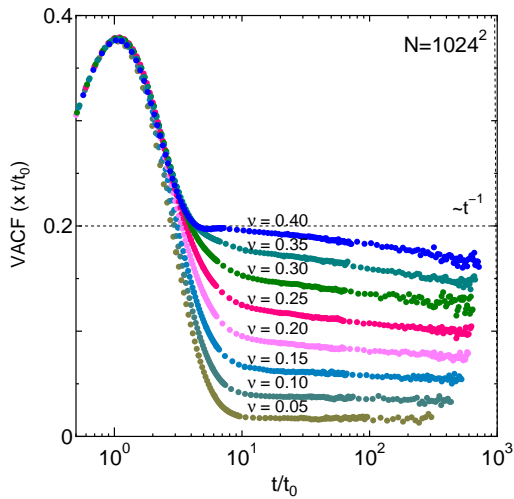}
\caption{
The packing fraction dependence of VACFs from a dilute to a moderately dense fluid with a one-million hard disk system is shown.
The vertical axis of the VACFs is multiplied by $t/t_0$ in the semi-log plot.
}
\end{figure} 

\begin{figure}
\includegraphics[scale=0.4]{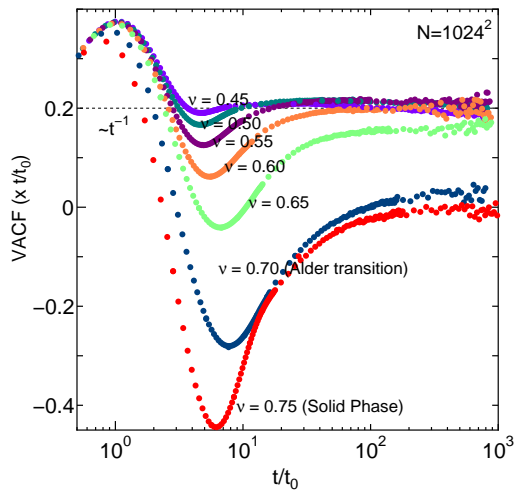}
\caption{
The packing fraction dependence of the VACFs from a moderately dense fluid to a solid with a one-million hard disk system is shown.
The vertical axis of the VACFs is multiplied by $t/t_0$ in the semi-log plot.
}
\end{figure} 

To investigate the functional form of the long time tail in detail, we show the semi-log plot of the VACFs for the packing fraction in Figs.~3 and 4.
The vertical axes are multiplied by $t/t_0$ to show the deviation from the conventional prediction of decay $(t/t_0)^{-1}$ more clearly.
In the stage of exponential decay $(t < \tau_E)$, the universal curves of the VACFs for each packing fraction have a maximum around $t/t_0  \sim 1$.
In the case of a dilute gas ($\nu=0.05$), the tail of the VACF ($t/t_0 > 10$) seems close to a flat line (the power decay with $\alpha \sim -1$).
By contrast, in a moderately dense gas ($\nu=0.15 \sim 0.35$), the decay of the VACFs is faster than the conventional power decay.
In Fig.~4, the ``back scattering'' effect gradually becomes more dominant than $\nu \sim 0.45$ around time $\tau_E$.
At $\nu=0.45 \sim 0.60$, this tendency becomes more remarkable.
In a dense fluid at $\nu=0.65$, VACF takes a negative value.
At $\nu=0.70 \sim 0.75$ (more than the Alder transition point), the effect of the solid-fluid transition causes a drastic change in the VACF, even when the decay is exponential.
We found that these results reveal new numerical results for the long time tail that show a peculiar density dependence.
These facts have never been investigated in previous works~\cite{alder_1970}, since the true VACF can only be seen within the time scale $t/t_0 \sim 30$.

From a theoretical perspective, another possibility for decay faster than the power form ($\sim (1/t)$) has already been proposed.
This is weak logarithmic divergence based on the self-consistent mode coupling theory (SCMCT)~\cite{kawasaki_1971, wainwright_1971}.
Ernst et al.~\cite{ernst_1970} predicted the VACF $C_D(t)$ based on MCT in a 2D hard disk fluid with the shear viscosity $\nu_{vis}$ as

\begin{equation}
C_D(t)\sim \frac{k_BT}{8m\pi(\nu_{vis}+D)}t^{-1}.
\end{equation}

\noindent
However, this solution results in a contradictory conclusion in assuming a finite value of the diffusion constant $D$ as $D$ diverges in the framework of linear response relations (Green-Kubo expression~\cite{kubo_1991}).
To avoid this problem, Kawasaki~\cite{kawasaki_1971} and Wainwright et al.~\cite{wainwright_1971} proposed the SCMCT independently, which is described by

\begin{equation}
C_D(t)\sim \sqrt{\frac{k_BT}{16m\pi}}\left(t\sqrt{\ln{(t)}}\right)^{-1}.
\end{equation}

\noindent
In this case, the diffusion coefficient $D(t)$ diverges weakly at the long time limit as $D \sim (\ln{(\infty)} + const.)^{\frac{1}{2}}$.
Wainwright et al.~\cite{wainwright_1971} stated that the two decays in Eqs.~(5) and (6) cannot be distinguished by their simulation within the maximum correlation time $t/t_0 \sim 30$.
\begin{figure}
\includegraphics[scale=0.4]{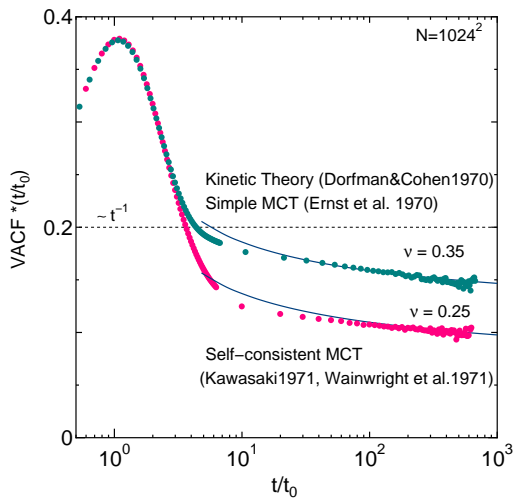}
\caption{
The VACFs in a moderately dense hard disk fluid are compared between numerical simulations and the theoretical predictions using the Simple MCT~\cite{ernst_1970}(broken line) and self-consistent MCT~\cite{kawasaki_1971,wainwright_1971} (solid line).
}
\end{figure} 
\noindent
Figure 5 compares numerical simulations of VACFs in a moderately dense fluid and the theoretical prediction of the SCMCT as a function in the form ($\beta \left(t\sqrt{\ln{(t)}}\right)^{-1}$) with a fitting parameter $\beta$.
Although the decay can be discussed within $t/t_0 \sim 10^3 $, the two functional forms at the long time limit seems to converge.


To summarize, we revisited the ``2D long-time-tail problem'' using a modern fast event-driven molecular dynamics simulation~\cite{isobe_1999}.
We completely reproduced the results of Alder and Wainwright, who discovered the power form of the long time tail~\cite{alder_1970}.
Compared with previous works, we essentially differ in having explored in detail a larger sample, longer correlation time, and more accurate VACFs.
We found that the tail of VACFs seem to decay as the power form ($\sim t^{-1}$) at the low density ($\nu=0.05$).
The remarkable results obtained from our simulation are the discovery of the decay of VACFs in a moderately dense hard disk fluid, which is faster than the conventional power decay.
We also compare the prediction of SCMCT~\cite{kawasaki_1971, wainwright_1971} seems consistent with our numerical results at a long time limit exceeding $t/t_0 \sim 100 $.
Although the coefficients of the prediction in the SCMCT change with the fitting parameter in Fig.~5, the tails seems to converge in the long time limit.
We conclude that a simple description of the decay of the power form over a long time limit might not be correct, at least in a moderately dense hard disk fluid.
Based on our results, further studies should reconsider 2D hard disk fluid using both extensive numerical simulations and the theoretical derivation of the transport coefficient (diffusion, viscosity, and heat conductivity) based on the linear response, kinetic theory, mode-coupling theory, and hydrodynamics.

I would like to thank Professor T. Y. Petrosky.
His theory of the long-time-tail problem based on the work of the Brussels school inspired this work.
I also acknowledge helpful comments made by Professors B. J. Alder, K. Kawasaki,  D. Frenkel and  H. Mori.
This work was supported by the Ministry of Education, Science, Sports and Culture, Grant-in-Aid for Scientific Research, Grant No.19740236.
A part of the computation in this work was done by the facilities of the Supercomputer Center, Institute for Solid State Physics, University of Tokyo.



\begin{thebibliography}{99}

\bibitem{alder_1970} B. J. Alder and T. E. Wainwright, Phys. Rev. A {\bf 1}, 18 (1970).

\bibitem{kubo_1991} R. Kubo, M. Toda and N. Hashitume, {\it Statistical Physics II} (Springer, 1991) Chap. 4.
; H. Nakano, Int. J. Mod. Phys. B {\bf 7}, 2397 (1993).


\bibitem{pomeau_1975} Y. Pomeau and P. R\'esibois, Phys. Rep. {\bf 19C}, 63 (1975).

\bibitem{dorfman_1977} J. R. Dorfman and H. van Beijeren, in {\it Modern Theoretical Chemistry Vol.6, Statistical Mechanics Part B}, edited by B.J.Berne (Plenum, New York, 1977), Chap. 3, p. 65.

\bibitem{dorfman_1970} J. R. Dorfman and E. G. D. Cohen, Phys. Rev. Lett. {\bf 25}, 1257 (1970).; Phys. Rev. A {\bf 6}, 776 (1972).

\bibitem{ernst_1970} M. H. Ernst, E. H. Hauge and J. M. J. van Leeuwen, Phys. Rev. Lett. {\bf 25}, 1254 (1970).

\bibitem{kawasaki_1970} K. Kawasaki, Phys. Lett. A {\bf 32}, 379 (1970).;Prog. Theor. Phys. {\bf 45}, 1691 (1971).;{\bf 46}, 1299 (1971).

\bibitem{petrosky_1999} T. Petrosky, Foundation of Physics {\bf 29}, 1417 (1999); {\bf 29}, 1581 (1999).

\bibitem{erpenbeck_1982} J. J. Erpenbeck and W. W. Wood, Phys. Rev. A {\bf 26}, 1648 (1982).; {\it ibid.} {\bf 32}, 412 (1985).; {\it ibid.} {\bf 43}, 4254 (1991).

\bibitem{frenkel_1989} D. Frenkel and M. H. Ernst, Phys. Rev. Lett. {\bf 63}, 2165 (1989).

\bibitem{hoef_1990} M. A. van der Hoef and D. Frenkel, Phys. Rev. A {\bf 41}, 4277 (1990).; Physica D {\bf 47}, 191 (1991).; Phys. Rev. Lett. {\bf 66}, 1591 (1991).

\bibitem{naitoh_1990} T. Naitoh, M. H. Ernst, and J. W. Dufty, Phys. Rev. A {\bf 42}, 7187 (1990).; T. Naitoh, M. H. Ernst, M. A. van der Hoef, and D. Frenkel {\bf 44}, 2484 (1991).

\bibitem{lowe_1995} C. P. Lowe and D. Frenkel, Physica A {\bf 220}, 251 (1995).

\bibitem{kawasaki_1971} K. Kawasaki, Phys. Lett. {34A}, 12 (1971).

\bibitem{wainwright_1971} T. E. Wainwright, B. J. Alder and D. M. Gass, Phys. Rev. A {\bf 4}, 233 (1971).

\bibitem{rapaport_1980} D. C. Rapaport, J. Comput. Phys. {\bf 34}, 184 (1980).

\bibitem{marin_1993} M. Mar\'in, D. Risso, and P. Cordero, J. Comput. Phys., {\bf 109}, 306 (1993).; M. Mar\'in and P. Cordero, Comput. Phys. Commun. {\bf 92}, 214 (1995).

\bibitem{isobe_1999} M. Isobe, Int. J. Mod. Phys. C {\bf 10}, 1281 (1999).

\bibitem{gaspard_2004} P. Gaspard and J. Lutsko, Phys. Rev. E {\bf 70}, 026306 (2004).

\end{thebibliography}

\end{document}